\documentclass[pre,onecolumn,floatfix,preprintnumbers,amsmath,
amssymb]{revtex4}

\usepackage{graphicx}
\usepackage{amsmath}
\usepackage{natbib}

\begin{document}
\title{Landau-Zener transitions in a semiconductor quantum dot}
\author{G.\ E.\ Murgida}
\author{D.\ A.\ Wisniacki}
\author{P.\ I.\ Tamborenea}
\affiliation{Departamento de F\'{\i}sica ``J.\ J.\ Giambiagi",
Universidad de Buenos Aires, Ciudad Universitaria, Pab.\ I,
C1428EHA Buenos Aires, Argentina}
\date{\today}

\begin{abstract}
We study the transitions between neighboring energy levels in a quasi-one-dimensional
semiconductor quantum dot with two interacting electrons in it, when it is subject to
a linearly time-dependent electric field.
We analyze the applicability of simple two-level Landau-Zener model to describe the evolution of the probability amplitudes in this realistic system.
We show that the Landau-Zener model works very well when it is viewed in the adibatic
basis, but it is not as robust in the diabatic basis. 

\end{abstract}

\maketitle

\section{Introduction}

The Landau-Zener (LZ) model describes in a simplified way the ubiquitous situation
of avoided crossings of energy levels in quantum mechanics \cite{zen}.
This happens in numerous areas of physics such as quantum optics, atomic
physics, nuclear physics, etc.
In spite of its simplified nature, the LZ model often captures the essential
features of avoided level crossings in realistic systems and the list of 
its applications continues to grow.

Recently, we have proposed a generic quantum control method based on the navigation 
of the energy spectrum.
The navigation of the spectrum is done varying a control parameter diabatically
and adiabatically.
The possibility of traveling through a complex spectrum depends crucially on
the nature of the energy level crossings.
Our method requires that the system behave locally (at avoided crossings) like
a LZ model, in the sense that complete diabatic and adiabatic transitions be 
possible.
So, the first step in the application of this method to a realistic system must
be a careful examination of the validity of this condition.
Note that in a realistic system the interaction between levels is often intricate
and the possibility that the LZ model worked has been discussed \cite{bul-dod-kus,
san-ver-wis}. 
Recently, we successfully applied this control strategy to a quantum dot system 
and the isomerization of a LiCN molecule.
In the present paper we analyse in detail the issue of the applicability of the
LZ model to the avoided crossings of the two-electron quantum-dot system studied
in Refs.\ \cite{mur-wis-tam,wis-mur-tam}.

Quantum dots are prime candidates to study the ideas and proposals of quantum control,
given their flexible and tunable properties.
In this paper, we continue the study of a quasi-one-dimensional double-dot system
with two interacting electrons.
Because of the one-dimensionality, this system is well suited to investigate
new methods of quantum control, and at the same time it incorporates the
important aspect of the interparticle interaction treated exactly.
We remark that the presence of interactions between particles is crucial in the
new science of quantum information processing.

The article is organized as follows: 
In order to make the work as self-contained as possible, in the next section  
we review the well-known Landau-Zener Model.
In Section \ref{sec:system} we describe our system and in the following section
we present the results concerning the applicability of the LZ model.
We conclude with some final remarks.

\section{Landau-Zener Model}
\label{sec:LZ_model}

\begin{figure}
\begin{center}
\includegraphics[height=7cm,angle=-90]{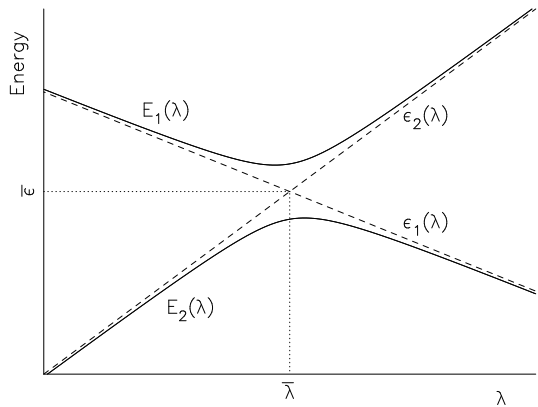}
\caption{Schematic plot of an avoided crossing.
The solid lines represent the adiabatic levels $E_1(\lambda)$ and $E_2(\lambda)$. 
The asymptotes to the energy hyperbolas are the diabatic straight 
lines $\varepsilon_1(\lambda)$ and $\varepsilon_2(\lambda)$ (dashed lines).
The center of the avoided crossing at $\overline{\lambda}$ and 
$\overline{\varepsilon}$ is inidicated with dotted lines.}
\label{fig:avoided}
\end{center}
\end{figure}

The LZ model \cite{zen} attempts to describe the universal situation of two levels 
interacting at an avoided crossing when a parameter $\lambda$ in the Hamiltonian 
is varied. The model consists of a two-level system described by
a parameter-dependent Hamiltonian, which expressed in the diabatic basis reads

\begin{equation}
 H=\left[\begin{array}{cc}
        \varepsilon_1(\lambda)  & \delta \\
        \delta                  & \varepsilon_2(\lambda)
     \end{array}\right],
\label{eq:hamiltonian}
\end{equation}
where  $\delta$ is a constant while $\varepsilon_1$ and $\varepsilon_2$ are linear 
functions of $\lambda$: 
$\varepsilon_1 = \overline{\varepsilon} + \alpha_1 (\lambda - \overline{\lambda})$, 
$\varepsilon_2 = \overline{\varepsilon} + \alpha_2 (\lambda - \overline{\lambda})$.
The center of the avoided crossing is located at $\overline{\lambda}$ and 
$\overline{\varepsilon}$ [see Fig.\ \ref{fig:avoided}].
The diabatic basis, $|1\rangle$ and $|2\rangle$, are parameter-independent 
eigenstates of the Hamiltonian Eq.\ (\ref{eq:hamiltonian}) with $\delta = 0$.

The eigenenergies  $E_1(\lambda)$ and $E_2(\lambda)$ of the Hamiltonian
(\ref{eq:hamiltonian}) are two hyperbolas (the adiabatic curves) as shown in 
Fig.\ \ref{fig:avoided}.
The eigenstates associated to those energies are the so-called adiabatic
states, which we denote $|\phi_1(\lambda)\rangle$ and 
$|\phi_2(\lambda)\rangle$.
The asymptotes to the energy hyperbolas are the diabatic straight 
lines $\varepsilon_1(\lambda)$ and $\varepsilon_2(\lambda)$. 
The shortest distance between the hyperbolas is $2 \delta$. 

In his seminal paper, Zener considered the parameter $\lambda$ as a linear
function of time and obtained the asymptotic probabilities of transitions
between the diabatic states in this time-dependent problem.
Assuming that the state $|1 \rangle $ is the initial state 
(at $t \rightarrow -\infty$) and $\lambda(t)= \beta \, t$, 
and calling $|\psi(t)\rangle$ the evolving wave function, the asymptotic 
probability to end up in the other diabatic state is  
\begin{eqnarray}
P_2(t \rightarrow \infty) &=& 
                 |\langle2|\psi(t \rightarrow \infty)\rangle|^2 = 
                 1 - \exp{\left[\frac{-2\pi \delta^2}
                            {\hbar |\dot{\varepsilon}_1-\dot{\varepsilon}_2|}
                 \right]} \nonumber \\
    &=& 1 - \exp{\left[\frac{-2\pi \delta^2}
                            {\hbar \beta |\alpha_1-\alpha_2|}
                 \right]}.
\label{eq:asymp_prob}             
\end{eqnarray}

The two limiting cases in terms of $\beta$ (the rate of change of the parameter 
$\lambda$) are: \\
i) Slow transition: 
$\beta \ll \frac{2 \pi \delta^2}{\hbar |\alpha_1-\alpha_2|}$. 
In this case the system follows the adiabatic curve going from the initial diabatic 
state to the other one. \\
ii) Rapid transition:
$\beta \gg \frac{2 \pi \delta^2}{\hbar |\alpha_1-\alpha_2|}$.
The evolution takes place on the diabatic curve and the system remains in the initial 
diabatic state.\\
These limiting cases play a central role in our control method \cite{mur-wis-tam}, 
as they give us quantitative criteria to choose either the diabatic
or adiabatic paths in traversing an avoided crossing \cite{wis-mur-tam}.

More recently, the complete time dependence of the occupation probabilities in both 
the diabatic ($|\langle 1 | \psi(t)\rangle|^2$, 
$|\langle 2 | \psi(t)\rangle|^2$)
and adiabatic basis  
($|\langle \phi_1(\lambda) | \psi(t)\rangle|^2, 
|\langle \phi_2(\lambda) | \psi(t)\rangle|^2$)
sets has been obtained \cite{vit,dam-zur}.


\section{The system: Quasi-one-dimensional doble quantum dot with two interacting 
electrons}
\label{sec:system}

Let us consider a quasi-one-dimensional double quantum dot with two interacting 
electrons in the presence of a uniform longitudinal electric field.
This system is interesting for two reasons. 
First, this type of system is experimentally realizable nowadays, and second,
the nonperturbative interparticle interaction is taken into account.

\begin{figure}
\begin{center}
\includegraphics[height=7cm,angle=-90]{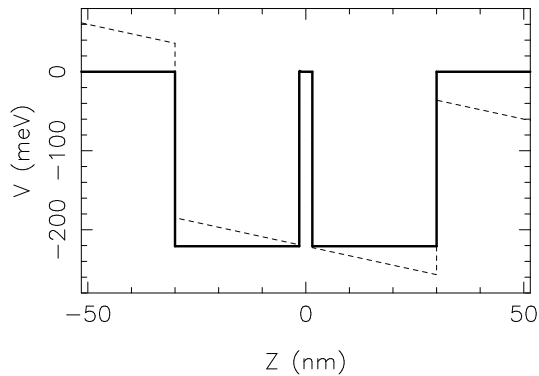}
\caption{Confining double well potential in the longitudinal direction of the coupled quantum dot structure. The external electric field is  $E=0$ (solid lines) and $E=12 \, \mbox{kV/cm}$ (dashed lines).}%
\label{fig:system}
\end{center}
\end{figure}

We have chosen a semiconductor system with realistic dimensions: it is an elongated
quantum dot $100 \, \mbox{nm}$ long and $50 \, \mbox{\AA}$ wide. 
Due to the small thickness of the structure, the energies of the transverse modes 
are widely spaced and it is enough to consider only the lowest transverse state.
Therefore, an effective Hamiltonian that depends only on the longitudinal coordinate
$z$ describes the dynamics of the system \cite{tam-met}
\begin{eqnarray}
\label{hamiltonian}
H&\equiv&-\frac{\hbar^2}{2m}(\frac{\partial^2}{\partial z_1^2} + 
\frac{\partial^2}{\partial z_2^2}) + V(z_1) + V(z_2) \nonumber
 \\
&&+ V_C(|z_1-z_2|) - e(z_1 + z_2)E(t) \, ,
\end{eqnarray}
where $m$ is the electron effective mass in the semiconductor material, 
$V_C$ is the Coulomb interaction between the electrons, 
$V$ is the confining potential, and 
$E(t)$ is a time-dependent external electric field.
There is no restriction regarding the choice of the confining potential in the 
$z$-direction, but we have selected the double-well configuration shown 
in  Fig.\ \ref{fig:system}. 
Double-well potentials are interesting due to the interplay between tunneling
and localization.
Moreover, in our two-electron system these important phenomena can be related
to the Coulomb interaction.
In all of the time evolutions that we will analyze, we assume that the wave function
is initially a singlet (antisymmetric spin wave function). 
Since the Hamiltonian is spin independent, the spin wave function remains a singlet
and the orbital part of the wave function is symmetric at all times.

We have used the time-dependent electric field $E(t)$ as the control parameter
\cite{mur-wis-tam,wis-mur-tam}. 
The first step is to understand the behavior of our system when the electric 
field is taken as a constant.
So, we have computed numerically the eigenergies and eigenfunctions of the system as a function of a constant electric field. 
We numerically diagonalize the Hamiltonian of Eq.\ (\ref{hamiltonian}) expanded  
in the basis set of Slater determinants constructed with the 12 bound single-particle
states of the double-well potential.
The two-particle basis set has then 12*(12+1)/2=78 states.

\begin{figure}
  \begin{center}
    \includegraphics[height=12cm,angle=-90]{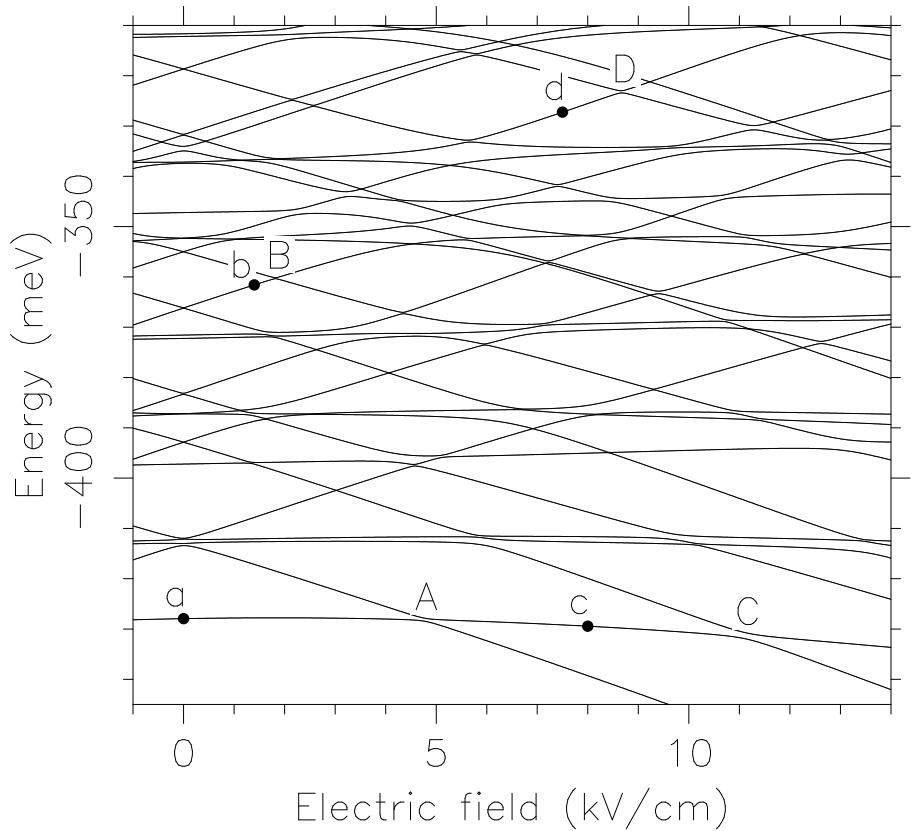}
    \caption{The energy spectrum of the two interacting electrons confined in
    a quasi-one-dimensional double-well semiconductor nanostructure as a function of 
    an external uniform electric field. It is plotted the first 31 energy levels.
    See text for details.}%
  \end{center}
  \label{fig:spectrum}
\end{figure}

The energy spectrum as a function of the external electric field is shown in 
Fig.\ \ref{fig:spectrum}. 
We can see that the spectrum is composed by fairly straight lines which never cross
each other, resulting in frequent avoided crossings.
For the low-lying states included in our spectrum, and far from the avoided crossings,
the adiabatic states have clear localization properties connected with their 
slope (see Fig.\ 2 in Ref.\ \cite{mur-wis-tam}): \\
i) in the eigenstates corresponding to negative slope both electrons are in the left 
well,\\ 
ii) the states associated to the positive slope are localized on the right well, and \\
iii) the states with neutral slope have one electron in each well.\\
At avoided crossings, the eigenstates mix their localization characteristics
reaching the maximal degree of mixing at the center of the avoided crossing.


\section{Analysis of the applicability of the Landau-Zener model}
\label{sec:analysis-applicability}

In our previous works we introduced a method of quantum control via traveling
in the energy spectrum of a quantum system \cite{mur-wis-tam,wis-mur-tam}.
The building blocks of this method are, on the one hand, the adiabatic evolutions
far from avoided crossings, and on the other, the slow  and fast 
evolutions employed at avoided crossings in order to shift in a controlled way
from one adiabatic path to a neighboring one.
If the system behaves locally like a LZ this possibility will be guaranteed.
For this reason, in this section we will analyze the range of validity or 
applicability of the 
Landau-Zener model to describe the transitions at the avoided crossings of 
our system.

\begin{figure}
  \begin{center}
    \includegraphics[height=10cm,angle=-90]{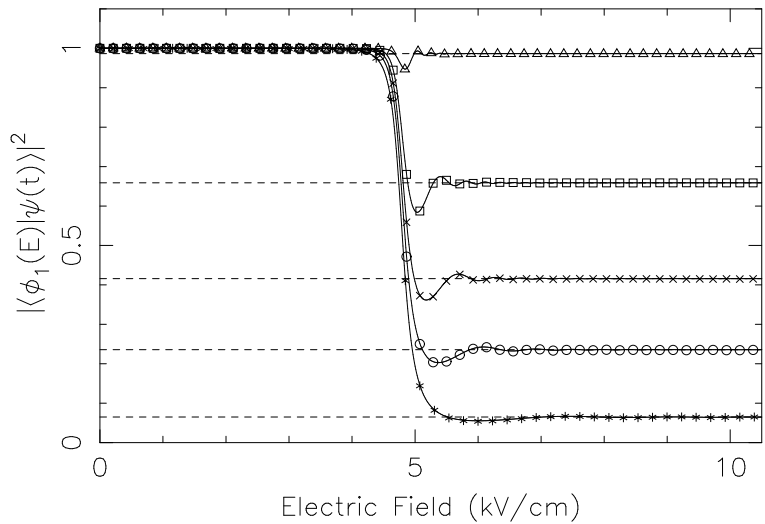}
  \end{center}
    \caption{Adiabatic transition probabilities 
             $|\langle \phi_1(E) | \psi(t)\rangle|^2$
             at the lowest avoided crossing (labelled ``A" in 
             Fig.\ \ref{fig:spectrum}) for various rates of change of 
             the control parameter.
              The velocities are: 
             $\dot{E} = 0.07 (\triangle), 0.27 (\Box), 0.53 (\times),     
             1.07 (\circ), 4.27 \,(*)\, \mbox{kV/cm ps}$.
             The initial state in the exact evolution is the one labelled ``a" 
             in Fig.\ \ref {fig:spectrum}.
             The solid lines give the adiabatic probabilities in the LZ model
             and the dotted lines are the asymptotic LZ probabilities
             given in Eq.\ (\ref{eq:asymp_prob}).}
 
  \label{fig:avoided_1_adiab}
\end{figure}

We begin by studying the avoided crossing labelled ``A" in Fig.\ \ref{fig:spectrum},
between the ground state and the first excited state near the value of the 
electric field $E = 5 \,\mbox{kV/cm}$.
Initially the system is in the ground state with no electric field 
(state labelled ``a" in Fig.\ \ref{fig:spectrum})  
and we study the probability to remain in the ground state when the electric field 
is increased linearly with time at different velocities.
This corresponds to the adiabatic probability 
$|\langle \phi_1(E) | \psi(t)\rangle|^2$
introduced at the end of Section \ref{sec:LZ_model}.
These probabilities are shown in Fig.\ \ref{fig:avoided_1_adiab}.
We remark that in order to compare the results for different velocities, we
plot in Fig.\ \ref{fig:avoided_1_adiab} the probabilities as functions of the electric field rather than as functions of time.
In Fig.\ \ref{fig:avoided_1_adiab} we present the adiabatic probability for the 
following velocities,
$\dot{E} = 0.07, 0.27, 0.53, 1.07, 4.27 \, \mbox{kV/cm ps}$.

We now compute the adiabatic probabilities in the LZ model.
The first step is to fit the parameters $\delta$, the location of the avoided crossing,
$\alpha_1$, and $\alpha_2$ \cite{footnote} 
of the two-level Hamiltonian of Eq.\ (\ref{eq:hamiltonian}) to the avoided crossing 
under study.
As initial state we take the adiabatic LZ state for the value of $\lambda$ that 
corresponds to state ``a" in Fig.\ \ref{fig:spectrum}.   
We compute the adiabatic probabilities in the LZ model for the previous set of rates
of change of the electric field. 
These results are plotted with solid lines in Fig.\ \ref{fig:avoided_1_adiab}.
We can see that our system is well described by the LZ model for the whole
range of velocities considered.
Moreover, we see that the asymptotic probability obtained by Zener (horizontal
dashed lines in Fig.\ \ref{fig:avoided_1_adiab}) gives accurate results far from the 
avoided crossing.
Note that far from the avoided crossing the diabatic and adiabatic states in the
LZ model are essentially the same.

\begin{figure}
  \begin{center}
    \includegraphics[height=14cm,angle=-90]{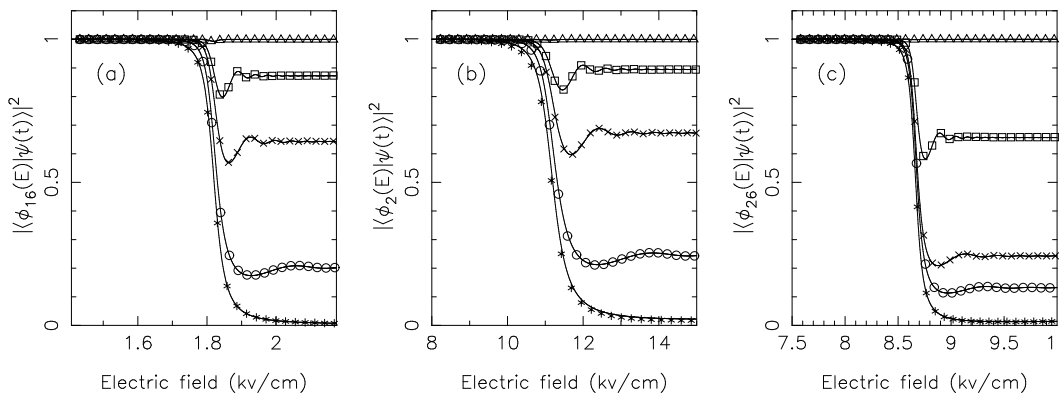}
   \end{center}
    \caption{Adiabatic transition probabilities 
             at various avoided crossings (labeled "B", "C" and "D" in 
             Fig.\ \ref{fig:spectrum}).
             The initial states in these evolutions are labelled in Fig.\ 
             \ref {fig:spectrum} as 
             "b" [panel (a)] , "c" [panel (b)], and "d" [panel (c)] . 
             The velocities are: 
             (a) $\dot{E} = 0.0015 (\triangle), 0.0077 (\Box), 0.015 (\times),     
             0.07 (\circ), 4.27 \,(*)\, \mbox{kV/cm ps}$, 
             (b) $\dot{E} = 0.07 \,(\triangle), 0.53 \,(\Box), 1.27 \,(\times),     
             4.27 \,(\circ), 40.0 \,(*)\, \mbox{kV/cm ps}$, and 
             (c) $\dot{E} = 0.003 \,(\triangle), 0.07 \,(\Box), 0.27 \,(\times),     
             0.53 \,(\circ), 4.27 \, (*)\, \mbox{kV/cm ps}$,}
  
  \label{fig:avoided_2_adiab}
\end{figure}

We have done similar analyses for other avoided crossings of the energy spectrum.
Namely, we start with the states labelled ``b", ``c", and ``d" in 
Fig.\ \ref{fig:spectrum}, and we study the transition probabilities in the 
adjacent avoided crossings ``B", ``C", and ``D", respectively.
In Fig.\ \ref{fig:avoided_2_adiab} we show the results, which verify the previous conclusion, in the sense that in the adiabatic basis the two-level LZ model fits 
very well the exact results.
It is worth noting here that in our previous work of 
Ref.\ \cite{mur-wis-tam,wis-mur-tam}
we travel in the spectrum (that is, we attempt to go from a given adiabatic
state to another one). 
In this sense, the above given results are the most relevant ones to judge
the applicability of our control method. 

\begin{figure}
  \begin{center}
    \includegraphics[height=10cm,angle=-90]{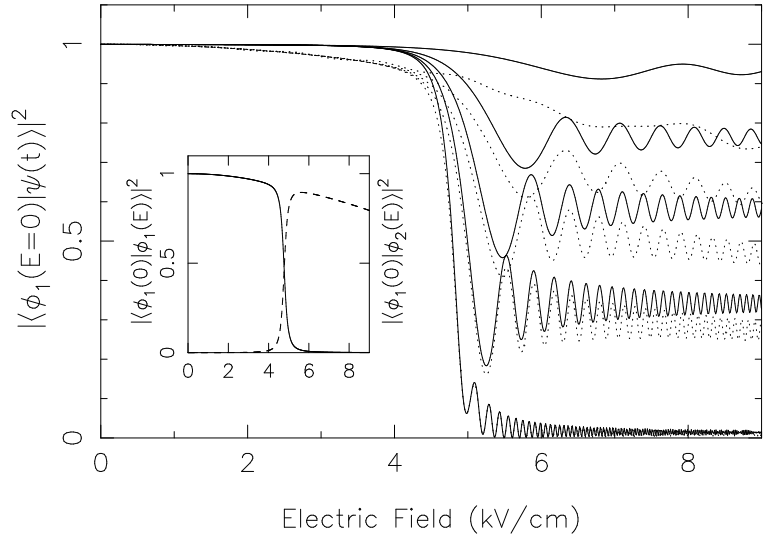}
    \end{center}
    \caption{Diabatic transition probabilities 
             $|\langle \phi_1(E_0=0) | \psi(t)\rangle|^2$ (dotted lines)
             at the lowest avoided crossing, labelled ``A" in Fig.\ \ref{fig:spectrum}. 
             The rates of change of the control parameter, the electric field $E$, are the
             same as Fig.\ \ref{fig:avoided_1_adiab}.
             In solid lines, the corresponding probabilities in the LZ model. 
             Inset: overlaps $|\langle \phi_1(E=0) | \phi_1(E) \rangle|^2$ (solid line)
             and $|\langle \phi_1(E=0) | \phi_2(E) \rangle|^2$ (dashed line).}
  \label{fig:avoided_1_diab}
\end{figure}

\begin{figure}
  \begin{center}
    \includegraphics[height=10cm,angle=-90]{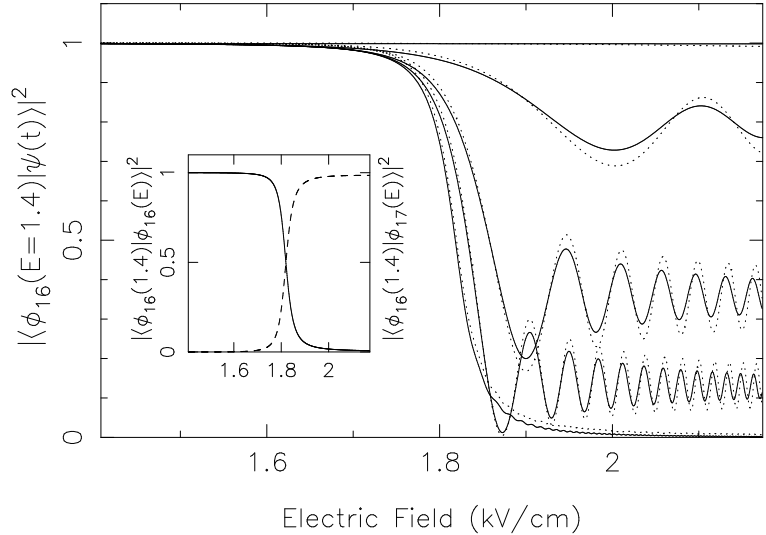}
    \end{center}
    \caption{Diabatic transition probabilities 
             $|\langle \phi_{16}(E=1.4 \, \mbox{kV/cm)} | \psi(t)\rangle|^2$ (dotted lines)
             at the avoided crossing, labelled ``B" in Fig.\ \ref{fig:spectrum}. 
             The rates of change of the control parameter, the electric field $E$, are the
             same as in Fig.\  \ref{fig:avoided_2_adiab}(a).
             In solid lines, the corresponding probabilities in the LZ model. 
             Inset: overlaps 
             $|\langle \phi_{16}(E=1.4 \, \mbox{kV/cm}) | \phi_{16}(E) \rangle |^2$ 
             (solid line)
             and $|\langle \phi_{16}(E=1.4 \, \mbox{kV/cm}) | \phi_{17}(E) \rangle |^2$
             (dashed line).}
  \label{fig:avoided_2_diab}
\end{figure}

Since the LZ model is defined on the basis of diabatic states, it is perhaps
more natural to perform the former analysis on that basis set.
However, the question arises of what the diabatic states are in our realistic
system.
Indeed, in a multilevel system like ours the two states involved in the
avoided crossing become mixed with other states and therefore acquire a 
dependence on the control parameter (which is not allowed for in the 
usual LZ model).
It is thus an interesting question to ask whether it is possible to find a 
``fixed" basis set which could play the role of the diabatic basis in 
the LZ model. 
For example, we now calculate the probability  
$|\langle \phi_1(E_0) | \psi(t)\rangle|^2$, where $|\phi_1(E_0)\rangle$ is the 
the initial state in the dynamic passage of an avoided crossing.
That is, we are considering $|\phi_1(E_0)\rangle$ as being one of the diabatic 
basis states. 
We now do this for the lowest crossing taking $E_0=0$, and compare with 
the results of using the two-level LZ model in Fig.\ \ref{fig:avoided_1_diab}.
One can clearly see that the agreement between the two calculations is 
not very good.
This can be understood with the help of the inset of 
Fig.\ \ref{fig:avoided_1_diab}, which shows
the overlaps $|\langle \phi_1(E=0) | \phi_1(E) \rangle|^2$ and
$|\langle \phi_1(E=0) | \phi_2(E) \rangle|^2$ as functions of
the electric field $E$.
It is clear from the inset, especially from  
$|\langle \phi_1(E=0) | \phi_2(E) \rangle|^2$ (dashed line), 
that the hypothesis of a parameter-independent diabatic state is not satisfied
(that the overlap is not equal to one far from the avoided crossing), 
and therefore the LZ model tends to fail.
However, in other avoided crossings we have observed that it is possible to find 
good diabatic states (which are fairly parameter-independent around the avoided crossing).
For example, we repeated the previous analysis for the avoided crossing labelled
``B" in Fig.\ \ref{fig:spectrum}, choosing 
$|\phi_{16}(E=1.4 \, \mbox{kV/cm})\rangle$ as one of the diabatic states.
In Fig.\ \ref{fig:avoided_2_diab}, we plot the probability 
$|\langle \phi_{16}(E=1.4 \, \mbox{kV/cm}) | \psi(t)\rangle|^2$, which shows
a better agreement than the one in Fig.\  \ref{fig:avoided_1_diab}.
We remark that, as can be seen in the inset of Fig.\ \ref{fig:avoided_2_diab}, the state 
$|\phi_{16}(E=1.4 \, \mbox{kV/cm})\rangle$ is a good choice of diabatic state,
since the overlap 
$|\langle \phi_{16}(E=1.4 \, \mbox{kV/cm}) | \phi_{17}(E) \rangle |^2$
is close to one at the right of the crossing and close to zero to the left
(see dashed line).
The behavior seen in the inset is exactly what one obtains in the LZ model
for the overlaps between the diabatic and adiabatic bases.

\section{Final Remarks}

We have studied the applicability of the LZ model in a realistic system:
a quasi-one-dimensional double quantum dot with two interacting electrons.
We showed that the LZ model works very well when it is viewed in the adiabatic
basis.
This result is the cornerstone for the quantum controlability using the
method of control introduced in \cite{mur-wis-tam}.

However, when seen in the diabatic basis the results are not so robust
as in the case of the adiabatic basis.
This is due to the fact that, for multilevel systems, a proper diabatic basis 
does not exist.
Rather, the pair of interacting levels at an avoided crossing become mixed with 
other states and acquire a dependence with the control parameter even far from
the avoided crossings.

\section*{Acknowledgement(s)}
The authors acknowledge support from CONICET (PIP-6137, PIP-5851) and
UBACyT (X248, X179).
D.A.W.\ and P.I.T.\ are researchers of CONICET.

\label{lastpage}


\begin{thebibliography}{99}
\bibitem{footnote}  Actually only the difference $\alpha_1 - \alpha_2$ enters in the
evolution of the two-level LZ system.  The location of the avoided crossing is irrelevant
in terms of the probabilities, but is is necessary to compare the results of the 
two calculations.
\end{thebibliography}
\end{document}